\documentclass[twocolumn,superscriptaddress,pre,floatfix]{revtex4-1} 

\usepackage{graphicx}
\usepackage{bm}	
\usepackage{color}
\usepackage{xcolor}
\usepackage{epsfig}
\usepackage{amsmath}
\usepackage{amssymb}
\usepackage{longtable}
\usepackage{float}
\usepackage{mathtools}
\usepackage{xparse}
\usepackage{times}
\usepackage{braket}
\usepackage[normalem]{ulem}

\begin{document}

\title{Decoherence of up to 8-qubit entangled states in a 16-qubit superconducting quantum processor}

\author{Asier~Ozaeta}
\email{asier.ozaeta@qcware.com}
\affiliation{QC Ware Corp., 125 University Ave., Suite 260, Palo Alto, CA 94301, USA}

\author{Peter~L.~McMahon}
\email{pmcmahon@stanford.edu}
\affiliation{E.\,L. Ginzton Laboratory, Stanford University, Stanford, California 94305, USA}
\affiliation{QC Ware Corp., 125 University Ave., Suite 260, Palo Alto, CA 94301, USA}
\date{\today}

\begin{abstract}
We report on the coherence of Greenberger-Horne-Zeilinger (GHZ) states comprised of up to 8 qubits in the IBM ibmqx5 16-qubit quantum processor. In particular, we evaluate the coherence of GHZ states with $N=1,\ldots,8$ qubits\footnote{In the cases of $N=1$ and $N=2$, one might formally call the states we studied GHZ-like, or generalizations of GHZ states, since a GHZ state is only defined for $N \ge 3$.}, as a function of a delay time between state creation and measurement. We find that the decay in coherence occurs at a rate that is linear in the number of qubits. This is consistent with a model in which the dominant noise affecting the system is uncorrelated across qubits.

\end{abstract}

\maketitle

\section{Introduction}

Prototype quantum processors based on superconducting qubits \cite{blais_quantum-information_2007,devoret_superconducting_2013}, and in particular transmon-related qubits, have become a leading platform for experimentally testing ideas in quantum information processing. The creation of a 10-qubit entangled state with transmon qubits has recently been demonstrated \cite{song_10-qubit_2017}; this followed earlier demonstrations of 3-qubit and 5-qubit entanglement \cite{dicarlo_preparation_2010,neeley_generation_2010,barends_superconducting_2014} with superconducting qubits. There are near-term plans to construct machines capable of entangling up to 49 qubits \cite{martinis_quantum_2017}, and it is expected that a processor with 49 qubits on a 2D lattice may, if the gate fidelities and qubit decoherence meet certain conditions, result in a demonstration of quantum computational supremacy \cite{boixo_characterizing_2016,harrow_quantum_2017}.

Studying the coherence of many-qubit entangled states can provide insight into the nature of the noise to which the qubits are exposed. The properties of the noise in a quantum computer can have profound implications for the error correction required to operate the computer fault-tolerantly \cite{knill_threshold_1996,aharonov_fault-tolerant_2006,aharonov_fault-tolerant_2008,reichardt_error-detection-based_2006,aliferis_level_2007,ng_fault-tolerant_2009,preskill_sufficient_2013,hutter_breakdown_2014,paz-silva_multiqubit_2017}. The strength and type of noise is also relevant in determining if a particular processor is performing a sampling task that is hard to simulate classically \cite{harrow_quantum_2017}.
GHZ states \cite{greenberger_going_1989,greenberger_multiparticle_1993} are canonical multi-particle entangled states in quantum information. They have been studied for their connection with the foundations of quantum mechanics and entanglement, but are also of interest in metrology \cite{toth_multipartite_2012,chaves_noisy_2013}. In this paper we study the decoherence of GHZ states in a 16-qubit superconducting processor; our choice of GHZ states is motivated both by the widespread appearance of GHZ states in the quantum information literature, and because a very similar study to the present one has been conducted with trapped-ion qubits, by Monz et al. \cite{monz_14-qubit_2011}, and we would like to allow easy comparison with their results.

\section{Methods}

We investigate the coherence of $N$-qubit Greenberger-Horne-Zeilinger (GHZ) states of the form $\ket{\psi}=\frac{1}{\sqrt{2}} \left( \ket{0 \ldots 0}+\ket{1 \ldots 1} \right)$ on the ibmqx5 16-qubit processor from IBM \cite{noauthor_ibmqx5:_2017}. We used a very similar procedure to evaluate the coherence to that described in Ref.~\cite{monz_14-qubit_2011}, which we outline below.

We perform, for each $N$, a set of experiments that together allows us to quantify how quickly the $N$-qubit GHZ state decoheres. The quantity we measure is the coherence $C$ of the GHZ state as a function of a delay time $\tau$ between state generation and a parity measurement that depends on the coherence. By the coherence $C(N,\tau)$ of the GHZ state, we mean specifically the following: the GHZ state under consideration can be represented by a density matrix $\rho^{(N,\tau)}$, and $C(N,\tau)$ is defined as the sum of the amplitudes of the far-off-diagonal elements $\rho^{(N,\tau)}_{11 \cdots 1,00 \cdots 0}$ and $\rho^{(N,\tau)}_{00 \cdots 0,11 \cdots 1}$, i.e., $C(N,\tau) \coloneqq \left| \rho^{(N,\tau)}_{11 \cdots 1,00 \cdots 0} \right| + \left| \rho^{(N,\tau)}_{00 \cdots 0,11 \cdots 1} \right|$.

In particular, for each $N \in \left\{ 1,2,\ldots,8 \right\}$, we run a circuit which itself has two parameters: a delay time $\tau$, and an analysis angle $\phi$. The circuit consists of four stages: (i) generate an $N$-qubit GHZ state of the form ($\ket{0 \dots 0}+\ket{1 \dots 1}/\sqrt{2}$);  (ii) introduce a delay $\tau$; (iii) rotate each qubit using the single-qubit unitary operator $U(\phi)$, and (iv) measure each qubit in the computational basis $\left\{ \ket{0}, \ket{1} \right\}$. This circuit is run with varying $\tau$, and for each $\tau$, it is run with $\phi$ ranging from $\phi=0$ to $\phi=\pi$, and for each combination of $N$, $\tau$, and $\phi$, the circuit is run multiple times to obtain sufficiently low statistical errors in the measurement results.

We aim to access information about the coherence $C(N,\tau)$ of the GHZ states; one convenient approach to measuring the coherence is to measure the amplitude of parity oscillations \cite{leibfried_creation_2005,monz_14-qubit_2011,sackett_experimental_2000}. Each qubit of a generated GHZ state is rotated by
\begin{equation}
U(\phi)=\cos\left({\frac{\pi}{4}}\right) I + i \sin\left({\frac{\pi}{4}}\right) \begin{pmatrix}0 & e^{-i\phi}\\ e^{i\phi} & 0\end{pmatrix}.
\end{equation}
These rotations induce oscillations in a measurable quantity called the parity $P \coloneqq P_\textrm{even} - P_
\textrm{odd}$ as the phase $\phi$ is varied. Here $P_\textrm{even/odd}$ correspond to the probabilities of finding the measured bitstring with an even/odd number of 1's. The amplitude of these oscillations is a direct measurement of the coherence $C(N,\tau)$ for a GHZ state with given number of qubits $N$ and a delay since generation $\tau$.

We investigate the coherence of each GHZ state as a function of time by varying the delay between creation and coherence measurement. In the experimental device under consideration, the observed coherence decay is exponential, and can be characterized by a coherence time parameter $T_2^{( N )}$, which we obtain by fitting an exponential function $\propto \exp\left( -t / T_2^{( N )} \right)$ to the observed $C(N,\tau)$ data.  This $T_2^{( N )}$ value is then compared with that of a single qubit; if the dominant noise source affecting the qubits is not correlated spatially, then one expects $T_2^{( N=1 )} / T_2^{( N )}=N$. This can be interpreted as the coherence time of an $N$-qubit GHZ state decreasing linearly with the number of qubits, or equivalently that the decoherence rate increases linearly with the number of qubits.

\begin{figure}
\centering
\includegraphics[width=0.5\textwidth]{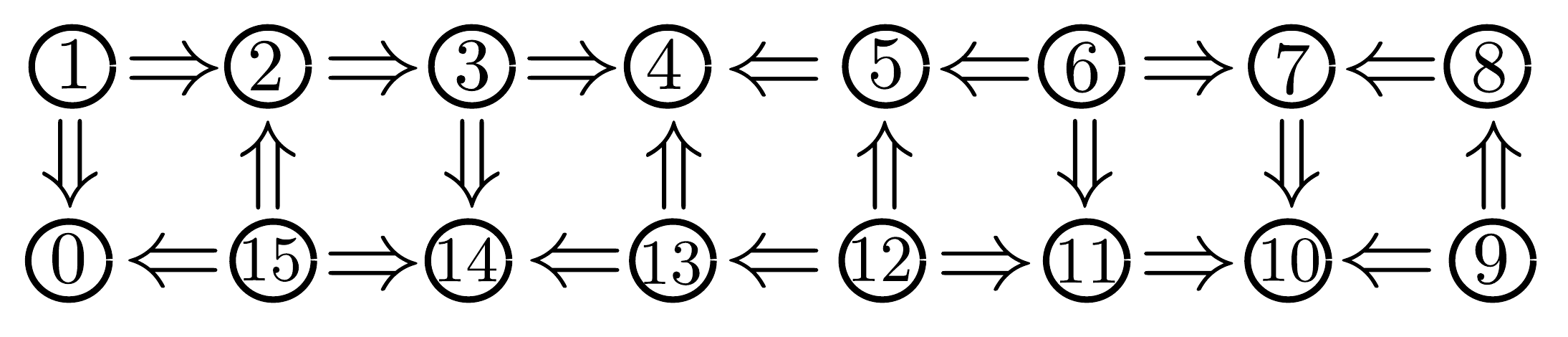}
\caption{\label{fig:connectivity}The layout of the ibmqx5 device, with the numbering of qubits used in this study. Lines show the direction in which CNOT gates are allowed between pairs of qubits, where $a \to b$ means that a CNOT gate with Qubit $a$ as control and Qubit $b$ as target can be performed.}
\end{figure}

We now describe some specifics about implementing the protocol described above on the ibmqx5 device. The ibmqx5 \cite{noauthor_ibmqx5:_2017} is a 16-qubit device in which every qubit can interact with at least two nearest neighbors via Controlled-NOT (CNOT) gates. The qubits are arranged in a $2 \times 8$ square lattice, with connectivity as shown in Fig.~\ref{fig:connectivity}. The connections between qubits have directionality: $a \to b$ means that only a CNOT with Qubit $a$ as control and Qubit $b$ as target is supported. To circumvent this limitation, one can apply Hadamard gates to the qubits acting as control and target before and after applying the CNOT gate in order to switch the direction.

This architecture allows for the generation of a 16-qubit GHZ state in principle. However, the finite gate fidelities in the device limit the size ($N$) of GHZ state that can be meaningfully prepared and analyzed in practice. In order to maximize initial fidelity we explored different gate paths to attempt to minimize the number of gates used to prepare the GHZ state. We used the QISKit SDK \cite{noauthor_qiskit-sdk-py:_2017} and the OpenQASM language for programming and running our circuits. 

\begin{figure}
\centering
\includegraphics[width=0.5\textwidth]{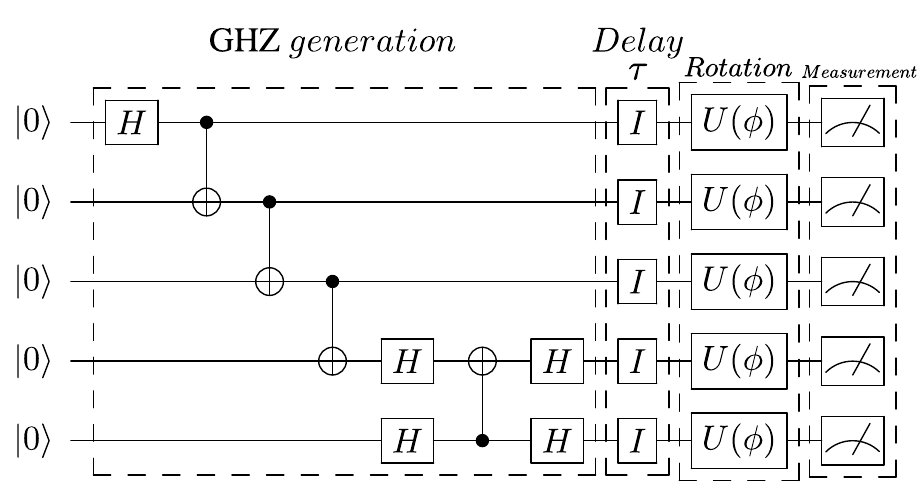}
\caption{\label{fig:circuit} Circuit used for the generation and analysis of GHZ states. The specific realization here is for the case $N=5$ qubits. The circuit shows only a single identity gate for simplicity; in practice the delay $\tau$ is implemented via the application of multiple fixed-time identity gates.
}
\end{figure}

Fig.~\ref{fig:circuit} shows the circuit we used for generating and analyzing the 5-qubit GHZ state. For the $N=5$ case, we used the physical qubits $\{1,2,3,4,5\}$, and owing to the directionality restrictions shown in Fig.~\ref{fig:connectivity}, this is a case in which the state generation required some additional gates beyond those in a canonical GHZ preparation circuit. The GHZ state is generate using a combination of Hadamard and CNOT gates. Because of the direction of the connection between the 4th and 5th physical qubits we are forced to include extra Hadamard gates to flip the control and target of the CNOT gate. A delay $\tau$ is realized via the application of multiple identity operations. Each identity operation has the same duration as a single-qubit rotation gate ($80~\textrm{ns}$) and is followed by a $10$-ns buffer time, which corresponds to a total delay of $90~\textrm{ns}$ per identity operation. The $U$ rotation needs to be implemented using the standard gates provided by the IBM API. We use the ibmqx5 standard unitary operator $U_3(\theta, \lambda,\phi_{U_3})$, which is defined as follows:
\begin{equation}
U_3(\theta, \lambda,\phi_{U_3}) \coloneqq \begin{pmatrix}  \cos{\frac{\theta}{2}} &  -e^{i \lambda} \sin{\frac{\theta}{2}}\\ e^{i\phi_{U_3}} \sin{\frac{\theta}{2}} & e^{i(\lambda+\phi_{U_3})} \cos{\frac{\theta}{2}}\end{pmatrix}.
\end{equation}
When $\theta=\pi/2$, $\phi_{U_3}=-\lambda$, and $\lambda=-\phi-\pi/2$, then $U_3(\theta, \lambda,\phi_{U_3}) = U(\phi)$, which is the desired rotation. Finally, we measure the qubits.

We choose which of the physical qubits of the available 16 to use as follows. For GHZ states with $N \in \left\{1,2,\ldots,6\right\}$, the qubits that comprise the GHZ states begin at qubit 1 and follow the numbering of the device up to $N$. For the $N=7,8$ GHZ states, we choose the physical qubits that maximize the coherence $C(N,\tau=0)$. Specifically, we used the chain $\left(4,13,12,11,10,9,8 \right)$ of physical qubits for $N=7$ and $\left(3,4,13,12,11,10,9,8 \right)$ for $N=8$.

\section{Results and Discussion}

\begin{figure}
\centering
\includegraphics[width=0.5\textwidth]{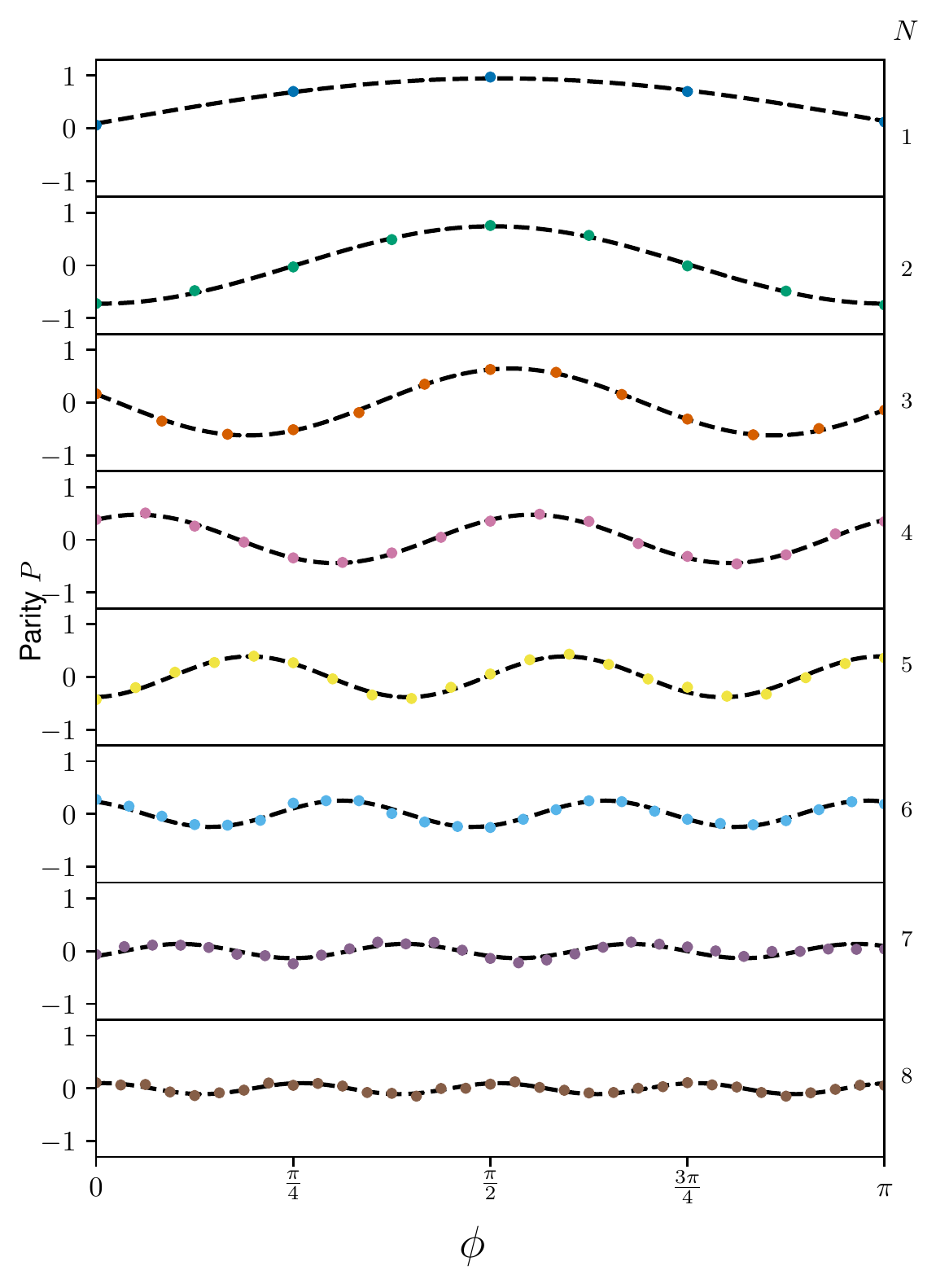}
\caption{\label{fig:subplot} 
Parity oscillations for each of the $N=1,\ldots,8$-qubit GHZ states generated on the ibmqx5, with delay $\tau = 0$. A sinusoid with fixed frequency, but free amplitude and phase variables, was fit to the data points for each $N$.
}
\end{figure}

Using the methods described above we experimentally measure the parity oscillations as a function of the phase of the rotation $\phi$, first with no delay ($\tau=0$) between generation and measurement. A sinusoid is fit to the data for each $N$. This is shown in Fig.~\ref{fig:subplot}. The amplitude of the fitted sinusoid corresponds to the coherence $C(N,0)$, since in this case the delay is zero. The coherence has a maximum value of $1$. Each point in the figure was obtained by performing $n=1000$ runs of the circuit, and an averaging of the results. In order to accurately fit the amplitudes of the parity oscillations, we sampled $4 N +1$ points for each $N$-qubit GHZ state. The period of the oscillations, $T(N)$, decreases with $N$ as $T(N)=2\pi/N$. We may get an estimation of the statistical error as a dispersion around the mean. We estimate the error of each point based on the mean parity $P_\textrm{even/odd}$ values and the number of data samples per point $n$ as $\delta P_\textrm{even/odd} \coloneqq \sqrt{P_\textrm{even/odd}(1-P_\textrm{even/odd})/n}$.

For every $N$, we expect that the parity $P$ will be equal to $0$ when $\phi=0$. However, for several values of $N$ we observed a shift in the oscillations, such that $P=0$ for values of $\phi \neq 0$.

\begin{figure}
\centering
\includegraphics[width=0.45\textwidth]{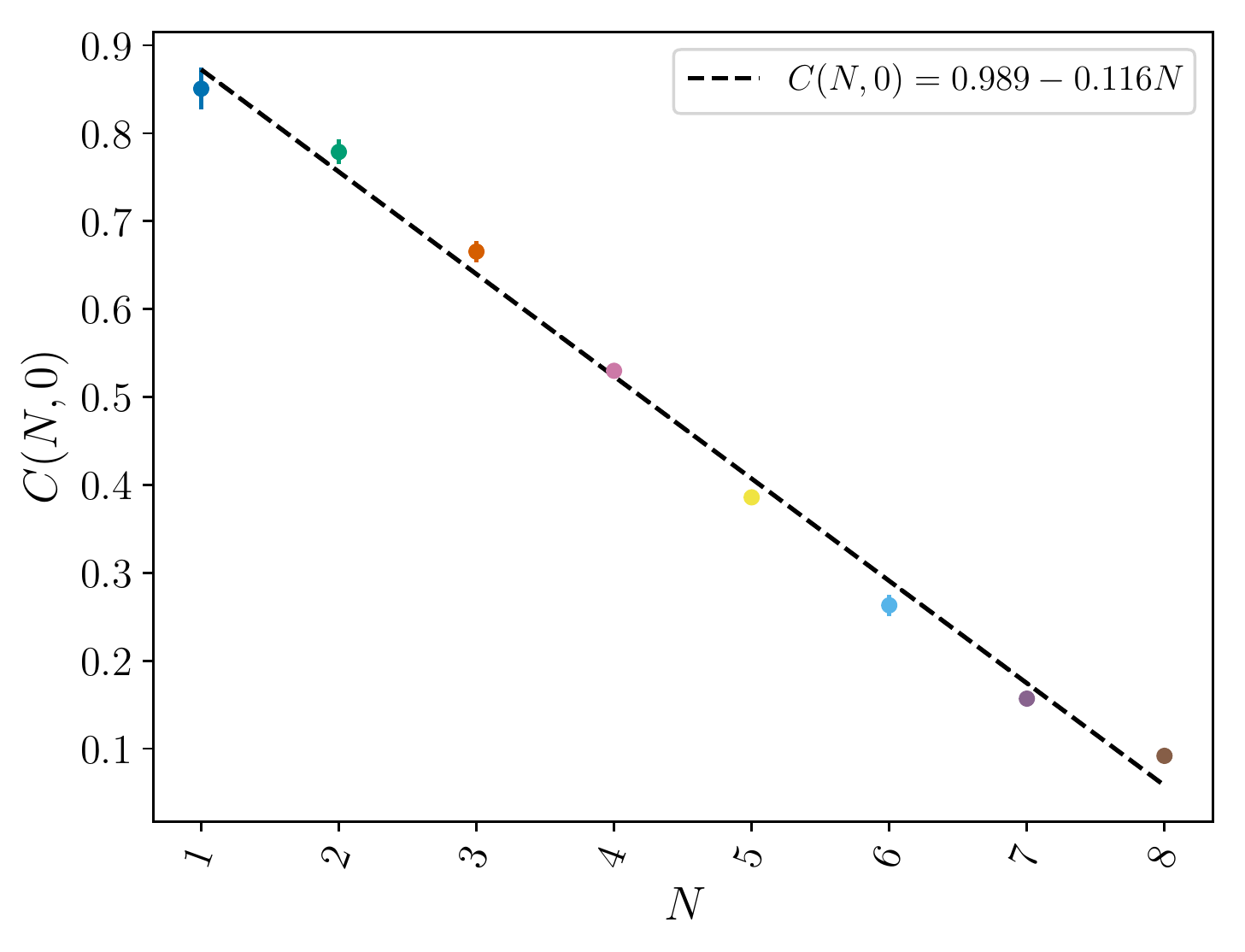}
\caption{\label{fig:amp_decay} Decay of the amplitude of the parity oscillations, which is the initial coherence $C(N,0)$, as a function of the size $N$ of each GHZ state. The dashed black line is a linear fit.}
\end{figure}

The coherence of each GHZ state with no delay $C(N,0)$, as obtained from the oscillations amplitudes in Fig.~ \ref{fig:subplot}, is plotted in Fig.~\ref{fig:amp_decay} as a function of $N$, and decreases approximately linearly with $N$ (we obtain a linear fit $C(N,0) \approx 0.88-0.12(N-1)=1-0.12 N$). Deviations from the linear trend can be partially attributed to the differences in fidelity for gates applied to different physical qubits, as well as to the fact that for some $N$, more than $N$ gates were used to generate the GHZ states due to the need to reverse the control and target qubits on some CNOT gates. We attribute the linear decrease of the initial coherence of the GHZ states to the linear number of quantum gates used to generate these states.

In this paper we studied GHZ states up to size $N=8$, even though the ibmqx5 chip has an architecture that could theoretically allow the generation of GHZ states as big as $N=16$. The reason for this is our inability to obtain clear parity oscillations for the $N=9$ case even after trying different combinations of physical qubits. The measured parity values no longer fit well to a sinusoid (see Appendix A), and so the initial (zero-delay) coherence $C(N=9,0)$ cannot be reliably measured. This inhibits meaningful assessment of how the coherence for states with $N \ge 9$ changes with added delay.

\begin{figure}
\centering
\includegraphics[width=0.5\textwidth]{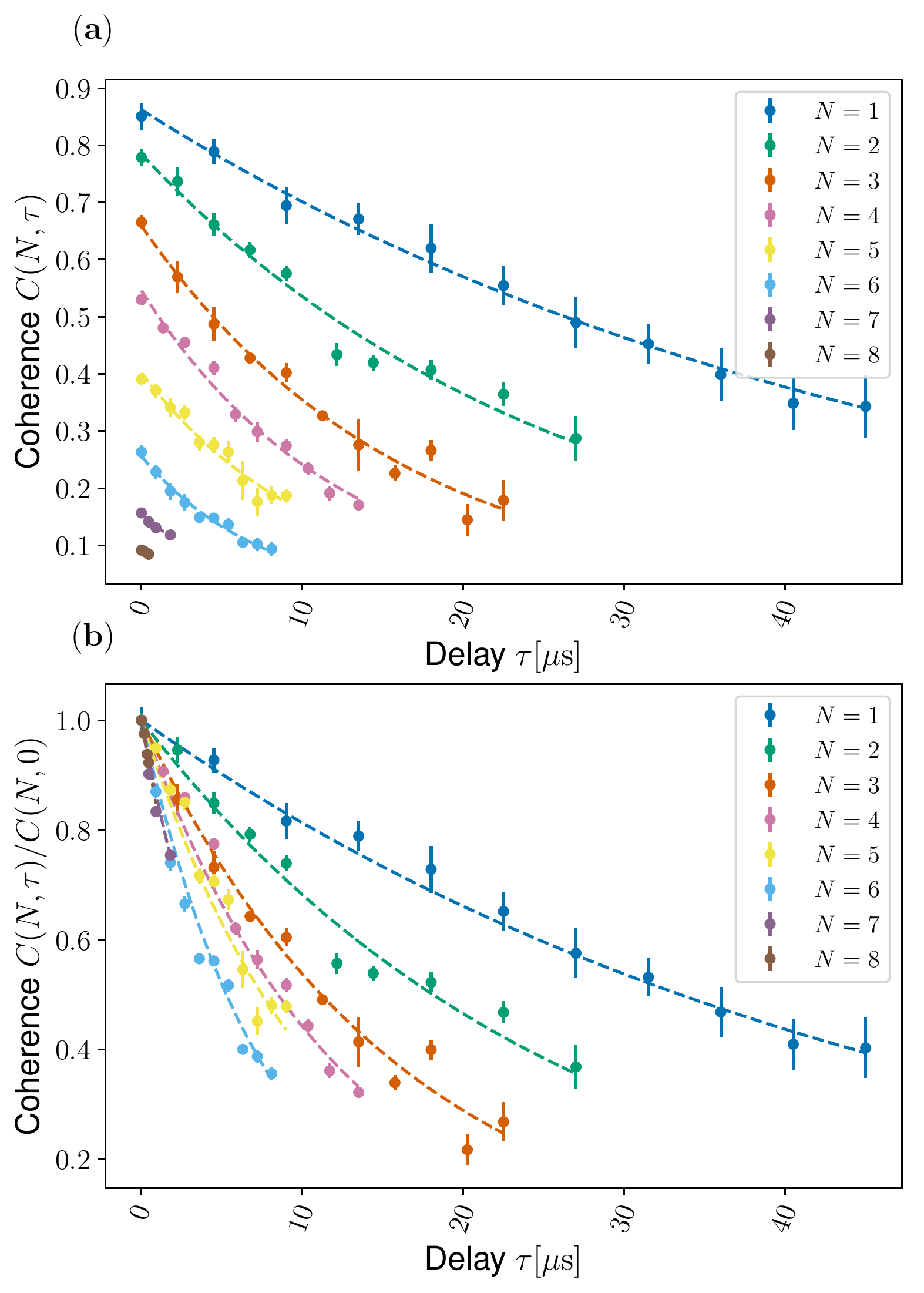}
\caption{\label{fig:exponential_decay} (a.) Coherence decay for GHZ states with different numbers of qubits as a function of the delay $\tau$. The points are from measured data and the dashed lines correspond to exponential fits for each $N$. The error bars correspond to the errors of the fits to the parity oscillations for each $N$ and $\tau$. (b.) Same as above, except now the coherence $C(N,\tau)$ is plotted as a normalized quantity with respect to the zero-delay coherence $C(N,0)$. This allows for easier visual comparison of the decay rates for each $N$.}
\end{figure}

For each $N=1,\ldots,8$, we measure how the coherence $C(N,\tau)$ reduces as a function of time by varying a delay $\tau$ between the generation and rotation of the GHZ states. The coherence reductions are manifest as reductions in measured parity oscillations as a function of $\tau$; for each $\tau$, we fit sinusoids to the parity oscillations and extract the fitted amplitudes, as we did for the case of $\tau = 0$ in Fig.~\ref{fig:subplot}. The decay of $C(N,\tau)$ as a function of $\tau$, for each $N$, is shown in Fig.~ \ref{fig:exponential_decay}(a). For each $N$, we fit an exponential decay function to the measured $C(N,\tau)$ data points: $c_N^\mathrm{init} \exp\left( -t/T_2^{(N)} \right)$, where $c_N^\mathrm{init}$ is the fit parameter that characterizes the initial GHZ state coherence, and $T_2^{(N)}$ is the fitted coherence time for the $N$-qubit state. Fig.~ \ref{fig:exponential_decay}(b) also shows the decay of coherence with $\tau$, but normalizes the coherence to $1$ at $\tau=0$ (normalizing out the difference in initial fidelity between the GHZ states of different size $N$), so that the monotonic increase in decay rates with $N$ can be seen more easily by eye.

The delay ranges that can be measured for the different number of qubits are limited either by our ability to measure clearly the coherence for a given $\tau$ or chosen based on our having measured sufficiently many data points to obtain a reliable $T_2^{(N)}$ fit ($N=1,2,3$).

\begin{figure}
\centering
\includegraphics[width=0.4\textwidth]{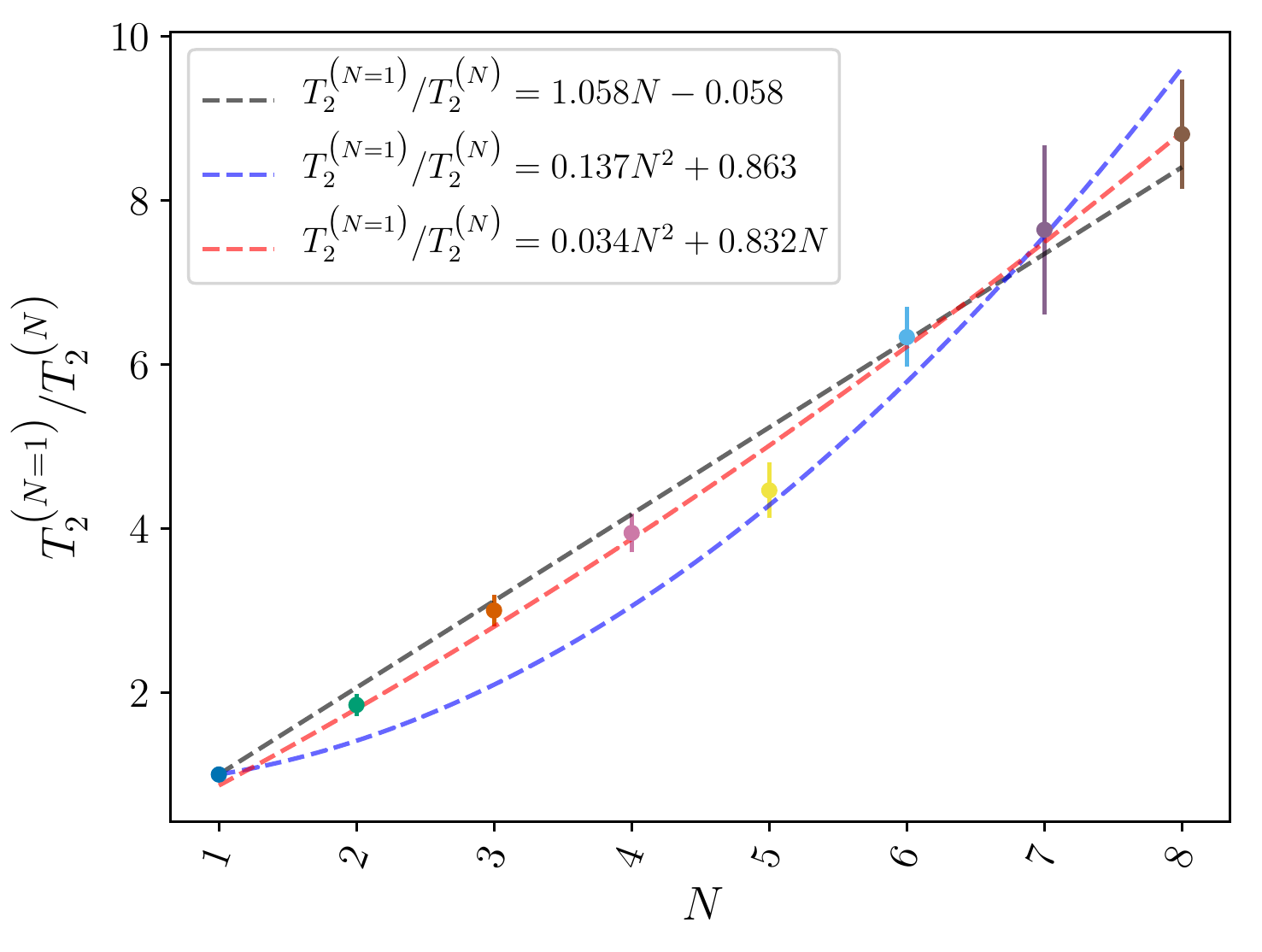}
\caption{\label{fig:ratio_error} The measured decoherence rate $1/T_2^{(N)}$ of each $N$-qubit GHZ state as a function of $N$, normalized by the decoherence rate for a single qubit, $1/T_2^{(N=1)}$.}
\end{figure}

The scaling of the $N$-qubit GHZ decoherence rate $1/T_2^{(N)}$, normalized by the single-qubit decoherence rate $1/T_2^{(N=1)}$, is shown as a function of $N$ in Fig.~\ref{fig:ratio_error}.
We have fitted three different functions of $N$ to the data: (i.) a linear function ($\beta N + \alpha$), (ii.) a quadratic function with the linear term set to zero ($\gamma N^2 + \alpha$), and (iii.) a quadratic function with the constant term set to zero ($\gamma N^2 + \beta N$). If the sources of decoherence for each qubit are independent, then the expected scaling in Fig.~\ref{fig:ratio_error} is linear. In contrast, if the system exhibits superdecoherence due to non-zero correlation in the noise, then a quadratic scaling may be expected. Fit (i.) has coefficient of determination $R^2 = 0.996$, and the $99\%$  confidence interval for the linear coefficient $\beta$ is $[0.968, 1.148]$. Fit (ii.) has $R^2 = 0.983$, and the $99\%$ confidence interval for the quadratic coefficient $\gamma$ is $[0.113, 0.160]$. Fit (iii.) has $R^2 = 0.998$, and the $99\%$ confidence intervals are $[0.561, 1.103]$ and $[-0.007, 0.075]$ for the linear ($\beta$) and quadratic ($\gamma$) coefficients respectively. While we cannot completely rule out that the true scaling of $T_2^{(N=1)}/T_2^{(N)}$ is non-linear, we note that the data presented in Fig.~\ref{fig:ratio_error} is well-fitted by a linear function, and the confidence interval of the slope is consistent with linear scaling $T_2^{(N=1)}/T_2^{(N)}=N$.

We note that Fig.~2(b) in Ref.~\cite{monz_14-qubit_2011}, which describes the increase in decoherence rate as a function of $N$ for a trapped-ion experiment, is directly comparable to Fig.~\ref{fig:ratio_error} in this paper, since the data were arrived at using nearly identical procedures.

\section{Conclusions}
In conclusion, we have analyzed the decay in coherence of GHZ states with up to 8 qubits in the ibmqx5 quantum computer. We find a linear increase in decoherence rate with the number of qubits, namely $T_2^{(N=1)}/T_2^{(N)}=N$.
The work of Monz et al. \cite{monz_14-qubit_2011} showed that GHZ states exhibited superdecoherence in a quantum processor comprised of $^{40}$Ca$^+$ ions, where each qubit was encoded in the electronic states $S_{1/2}$ and $D_{5/2}$ of a single ion. Since these states are not insensitive to magnetic fields, fluctuations in the current in Helmholtz coils (which form part of the experimental apparatus) lead to correlated dephasing noise for all the qubits. The use of magnetic-field insensitive (``clock'') states, or decoherence-free subspaces, can be used to mitigate the deleterious effects of magnetic-field noise in trapped-ion processors \cite{langer_long-lived_2005}.  However, our results provide evidence that superconducting processors constructed from single-junction transmons do not need further engineering to avoid superdecoherence (at least in the design and scale of the ibmqx5 system). This conclusion is consistent with current understanding of the known dominant sources of decoherence for IBM's transmon qubits \cite{gambetta_building_2017}, which are thought to act independently on each qubit.

\section{Acknowledgments}
We gratefully acknowledge the IBM-Q team for providing us with access to their 16-qubit platform. The views expressed are those of the authors and do not reflect the official policy or position of IBM or the IBM Quantum Experience team. We thank T.~Monz and J.\,I.~Adame for helpful discussions, and J.\,I.~Adame for a thorough reading of a draft of this paper.

\appendix

\section{9-qubit GHZ-state generation}

\begin{figure}
\centering
\includegraphics[width=0.4\textwidth]{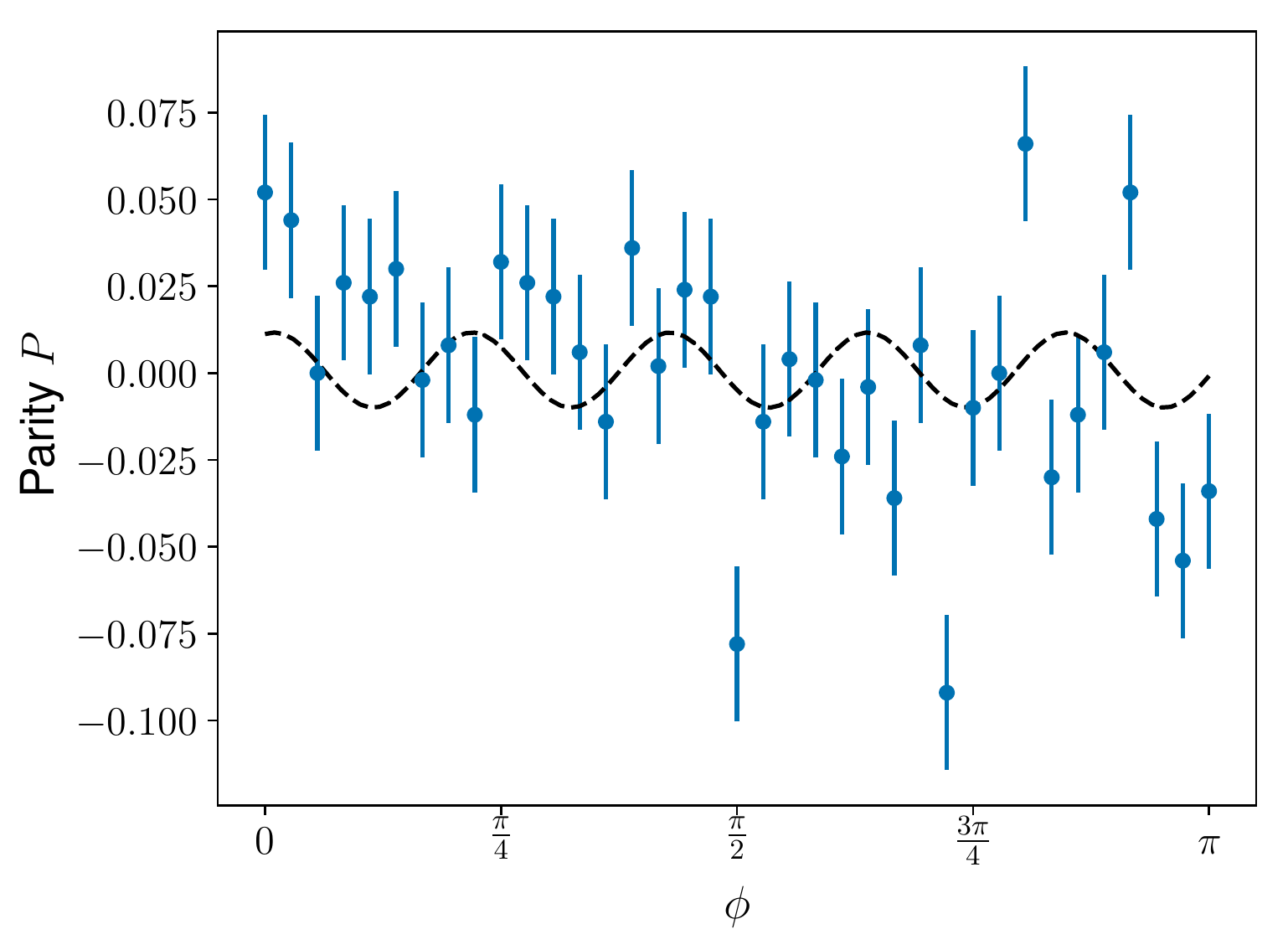}
\caption{\label{fig:osci_9} Parity measurement as a function of $\phi$ for the $N=9$ GHZ state. An attempt to fit a sinusoid to the measured data doesn't yield meaningful amplitude or phase parameters because a single-frequency oscillation is not apparent in the data. This is indicative of very low fidelity of the created state.}
\end{figure}

\begin{figure}
\centering
\includegraphics[width=0.47\textwidth]{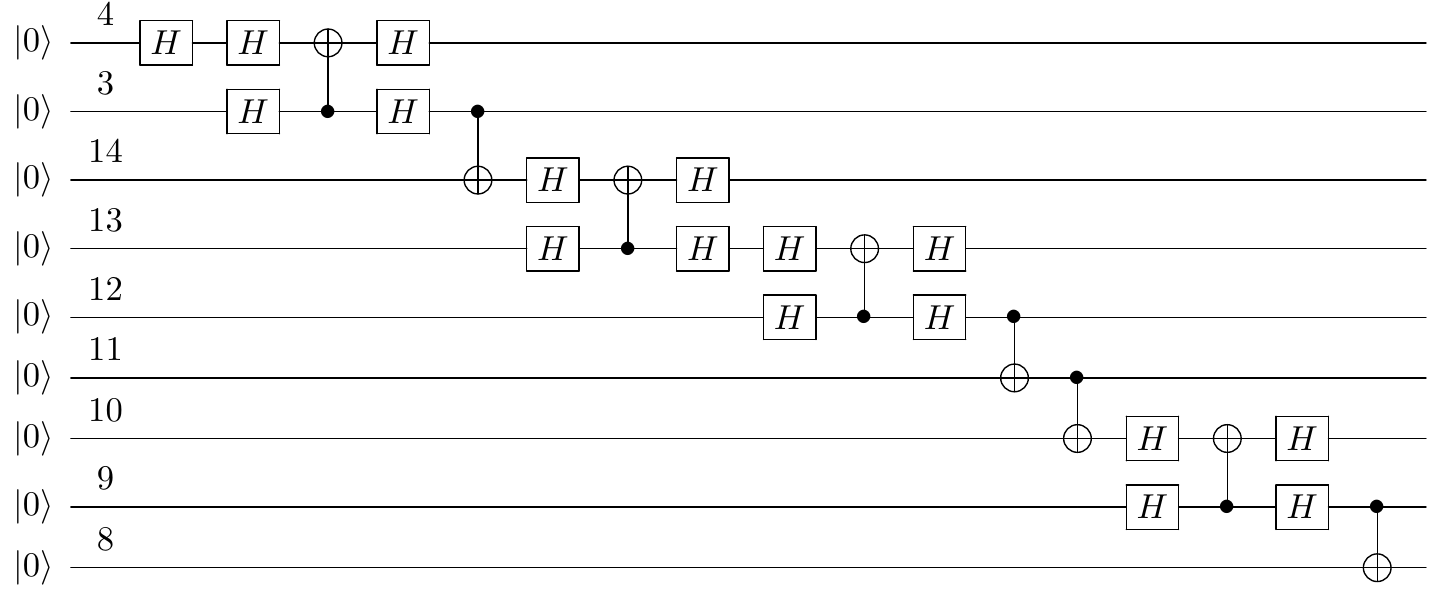}
\caption{\label{fig:circuit_9} Circuit used for the realization of the $N=9$-qubit case. The delay, rotation, and measurement parts are not displayed. The labels correspond to the physical qubit numbers. }
\end{figure}

In this paper we study GHZ states of size $1 \leq N \leq 8$, even though the ibmqx5 chip contains 16 qubits and a connectivity between them that in principle allows generation of GHZ states with $N$ up to 16. The reason for this is our inability to obtain clear parity oscillations for the $N=9$ case, despite attempts we made using different combinations of physical qubits.  Fig.~\ref{fig:osci_9} shows the oscillations that we observe for zero delay using the physical qubits $\{4,3,14,13,12,11,10,9,8 \}$, and the corresponding sinusoidal fit. The points do not appear to follow a sinusoidal oscillation as a function of $\phi$. The amplitude of the parity is also considerably reduced with respect to the $1 \leq N \leq 8$ cases. The circuit used to implement the $N=9$ GHZ state is shown in Fig.~\ref{fig:circuit_9}. We attribute our failure to observe parity oscillations for the $N=9$ GHZ state to the ibmqx5 chip having insufficient quantum volume \cite{quantum_volume}.

\section{Fitted $T_2^{(N)}$ values}

Table~I lists the specific numerical values of $T_2^{(N)}$ that were obtained in our experiments. These values characterize the coherence times of $N$-qubit GHZ states in the processor.

\begin{table}[H]
\label{tab:FittedT2N}
\begin{center}
 \begin{tabular}{||c | c ||} 
 \hline
 $N$ & $T_2^{(N)}$ $[\mu \textrm{s}]$ \\ 
 \hline
  1 & 48.34 $\pm$ 1.56 \\
  2 & 26.15 $\pm$ 1.67 \\
  3 & 16.11 $\pm$ 0.89 \\
  4 & 12.25 $\pm$ 0.62 \\ 
  5 & 10.83 $\pm$ 0.75 \\
  6 & 7.63 $\pm$ 0.36 \\
  7 & 6.32 $\pm$ 0.83 \\
  8 & 5.49 $\pm$ 0.38 \\ 
 \hline
\end{tabular}
\end{center}
\caption{Fitted values of $T_2^{(N)}$ (the coherence times of $N$-qubit GHZ states) from the data in Fig.~\ref{fig:exponential_decay}.}
\end{table}

\section{Theoretical evaluation of expected $T_2^{(N)}$ values}

Table~II lists the theoretical expectation of $T_2^{(N)}$ based on detailed calibration data from the IBM quantum experience at the time of the measurements. We obtain $T_2^{(N)}$ from the coherence times of the individual qubits. These values are comparable to the fitted values of the previous section. 

Assuming that errors are uncorrelated, we take into consideration $T_1$, $T_2$, gate errors, and readout errors to calculate the initial coherence. The coherence values show that the requirement for more gates to generate the GHZ state causes an increased time for the system to decohere.

\begin{table}[H]
\label{tab:CalcT2N}
\begin{center}
 \begin{tabular}{||c | c ||} 
 \hline
 $N$ & $T_2^{(N)}$ $[\mu \textrm{s}]$ \\ 
 \hline
  1 & 44.4 \\
  2 & 24.52 \\
  3 & 17.21 \\
  4 & 14.75 \\ 
  5 & 10.97 \\
  6 & 9.88 \\
  7 & 5.99 \\
  8 & 5.40 \\ 
 \hline
\end{tabular}
\end{center}
\caption{Calculated values of $T_2^{(N)}$, obtained from the calibration data.}
\end{table}

\section{Log-scale plots}
\begin{figure}
\centering
\includegraphics[width=0.5\textwidth]{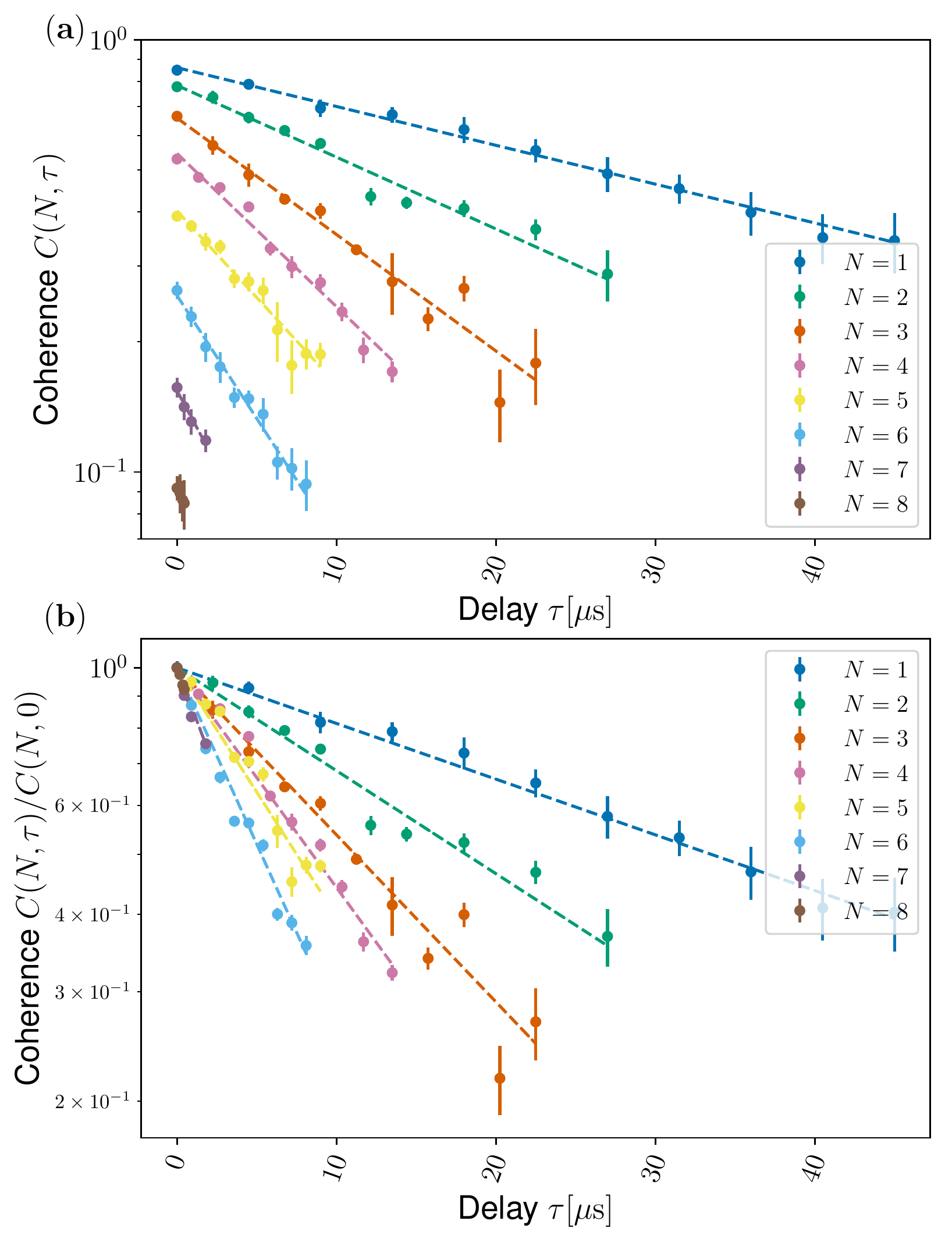}
\caption{\label{fig:exponential_decay_log}Re-plotting of the data of Fig.~\ref{fig:exponential_decay} on a log scale. }
\end{figure}

Fig.~\ref{fig:exponential_decay_log} shows a re-plotting of Fig.~\ref{fig:exponential_decay} on a log scale; this allows for easier visual inspection that the decays are indeed exponential.

\bibliographystyle{h-physrev}
\bibliography{PLM}

\end{document}